\documentclass[a4paper,fleqn,usenatbib]{mnras}

\usepackage{newtxtext,newtxmath}

\usepackage{ae,aecompl}

\usepackage{natbib}
\usepackage{amssymb}
\usepackage{graphicx}
\usepackage{bigfoot}
\let\ni=\noindent

\title[Topology of large scale under-dense regions]
      {Topology of large scale under-dense regions}

\author[A. M. So\l tan]{A. M. So\l tan\thanks{E-mail:
soltan@camk.edu.pl}\\
Nicolaus Copernicus Astronomical Centre, Polish Academy of Science,
Bartycka 18, 00-716 Warsaw, Poland }

\begin{document}

\date{Accepted \hspace{15mm}. Received  \hspace{15mm}; in original form }

\pagerange{\pageref{firstpage}--\pageref{lastpage}} \pubyear{}

\maketitle

\label{firstpage}

\begin{abstract} 
We investigate the large scale matter distribution adopting QSOs as matter
tracer. The quasar catalogue based on the SDSS DR7 is used. The void finding
algorithm is presented and statistical properties of void sizes and shapes are
determined. Number of large voids in the quasar distribution is greater than
the number of the same size voids found in the random distribution.  The
largest voids with diameters exceeding $300$\,Mpc indicate an existence of
comparable size areas of lower than the average matter density. No void-void
space correlations have been detected, and no larger scale deviations from the
uniform distribution are revealed. The average CMB temperature in the
directions of the largest voids is lower than in the surrounding areas by
$0.0046 \pm 0.0028$\,mK. This figure is compared to the amplitude of the
expected temperature depletion caused by the Integrated Sachs-Wolfe effect.
\end{abstract}


\section{Introduction}

Statistical characteristics of matter distribution depend on a number of
cosmological parameters. Albeit structures on various scales carry the
information of cosmological relevance, matter agglomerations on the largest
scales attract the greatest interest. This is simply because the large
structures are rare, and partly because some `unusual' accumulations of
matter could impose unique constraints on the precision cosmology
$\Lambda$CDM model.  A question of identifying structures is of statistical
nature and has a long history. It was recently discussed by \citet{park15}.
Here we examine statistics of large voids found in the SDSS DR7 quasar
catalogue.

Space distributions of individual matter components -- luminous matter,
diffuse baryonic matter and dark matter -- are strongly correlated, but not
identical \citep[e.g.][]{suto04}.  Also individual types of galaxies do not
follow one universal distribution pattern. It seems, however, that
noticeable differences that show up at small scales, systematically
disappear at large scales. In particular, it seems legitimate to assume
that at scales of hundreds Mpc the distribution of baryonic matter follows
that of the dark matter.

Quasars are suitable to study the matter distribution at the largest scales
for several reasons. Being the most luminous active galactic nuclei,
samples of quasars cover usually huge volumes. Magnitude limited samples of
quasars show lower than normal galaxies radial density gradients because of
strong cosmic evolution. This allows to construct voluminous data sets with
low observational selection bias \citep[e.g.][see below]{croom01}.
Clustering properties of quasars and galaxies are not distinctly different
at small and medium scales \citep[see][]{ross09}, what assures us that at
scales of several hundreds Mpc quasar distribution is representative for
the luminous matter distribution. To be more specific, the relationship
between the spatial distribution of galaxies and quasars is expected to be
linear, i.e.  amplitudes of the relative fluctuations of both components
are equal.

Numerous galaxy surveys reveal a variety of structures that span a very
wide range of linear sizes. Early 3D maps show pronounced filaments and
voids extending over $50$ and more Mpc
\citep{tarenghi79,chincarini83,huchra83, delapparent86}. Still larger
structures based on the SDSS have been reported; in particular, the Sloan
Great Wall $420$\,Mpc long \citep{gott05}, and the largest filamentary
structure extending above $1$\,Gpc found in the DR7QSO catalogue
\citep{clowes13}. However, statistical significance of the latter one was
questioned by \citet{park15} on the grounds that group finding algorithm
used by \citet{clowes13} was not sufficiently restrictive and structures
formed by chance were recognized as physical quasar group.

A question of distinction between `real' and `by chance' structures is
crucial for statistical studies of the largest matter accumulations
observed in the Universe. Long filaments may arise as a result of a
coherent process that involves simultaneously adequately big amount of
matter, or may be a product of chance alignment of separate `short'
filaments \citep{park15}. Likewise, large volumes of low matter density may
develop from a single large scale fluctuation, or be a close group of
`normal' voids similar to those observed in the local Universe.  In the
present paper we address this last question. Distribution of quasars in the
SDSS DR7 catalogue is investigated in respect of the number and size of
empty regions (voids). Then, shapes of voids and void correlation is
examined.

All distances and linear dimensions are expressed in co-moving coordinates.
To convert redshifts to the co-moving distances, we use the flat
cosmological model with $H_{\rm o} = 70$\,km\,s$^{-1}$Mpc$^{-1}$,
$\Omega_{\rm m} = 0.30$ and $\Omega_{\Lambda} = 0.70$. We focus our study
on the area between $3000$ and $4500$\,Mpc, what corresponds approx. to the
redshift range $0.8 - 1.6$.

The present investigation seems to belong to a broad field of galaxy
distribution studies that adopt a void concept. However, our work is rather
weakly related to this area. This is because of several reason. The most
conspicuous are the void scale and definition. Here we examine voids with
radii above $145$\,Mpc that are completely empty.  i.e. with no objects
inside.  Such zero-one approach is usually dropped in the galaxy studies.
In our geometrical attitude to void definition we ignore kinematic effects
of deviations from the Hubble flow. The redshift -- distance relationship
is determined by the cosmological model. Also, questions on dynamics of the
large scale inhomogeneities  of the  matter distribution cannot be
addressed here at the present stage.

Concentration of quasars is several orders of magnitude lower than the
galaxy space density. Consequently, the area nominally covered by the DR7
QSO catalogue is sparsely populated with the average distance between
neighbouring quasars of $r_{\rm n} \sim 125$\,Mpc. Statistical relationship
between the galaxy and quasar space distribution holds for scales
considerably larger than $r_{\rm n}$, and only  voids covering several
$r_{\rm n}$ indicate areas of the actually low galaxy density and the low
total matter density\footnote{One should note that the number of large
voids depends strongly on the local average density of points (see
Fig.~\ref{fig:sim_vd_number} in the Appendix~\ref{app:poisson}).
Obviously, this effect depends on the cosmic fluctuations of quasar
concentration, as well as on depth and homogeneity of the survey.}.
Because of that, our analysis of quasar voids properties is limited to the
largest voids.  An advanced  method to investigate topology of continuous
fields in cosmology, namely a watershed void finder (WVF), has been
developed in recent years \citep*{platen07}. WVF is a particularly
effective tool to study evolution of complex, hierarchical structures both
in the real data and simulations \citep[][and references
therein]{vandeweygaert16}. The observational material is practically always
represented by the discrete samples, and\ transformation of the discrete
distribution into the continuous one constitutes the inherent element of
the WVF method \citep{schaap00}. Here we will examine scales only a few
times larger than the average quasar separations (see below).  Voids in the
present situation are defined by just a few objects, what does not allow
for a credible construction of the continuous density field.

N-body simulations show that in scales of few dozens MPC void
properties evolve with time \citep{sheth04}.  Because the present
investigation concentrates on much larger structures than typical voids
observed in galactic catalogues, and we concentrate on a single large
volume contained within a relatively narrow range of redshifts ($0.8
\lesssim z \lesssim 1.6$), no cosmic evolution effects are considered in
the paper.

Voids detected in galaxy surveys span a wide range of sizes. For instance,
\citet{lares17} identify voids in the SDSS DR7 galaxy catalogue with radii
in the range $6 - 24\,h^{-1}$\,Mpc.  Only relatively small and moderate
size voids are almost completely devoid of galaxies
\citep{vandeweygaert16}.  Linear sizes of the present voids exceed
$300$\,Mpc, and are comparable to the largest voids detected in the galaxy
distribution \citep{stavrev00}. Such large volumes are called voids because
of distinctly lower than average number of galaxies.  However, the
amplitude of density fluctuation within large voids is not well determined
\citep{kopylov02}. Most information on this question is obtained from
N-body large scale simulations \citep{nadathur17}. Also the present
investigation cannot provide direct information on the distribution of
galaxies in the areas coinciding with the quasar voids. One of the
objectives of our paper is to measure the amplitude of the Integrated
Sachs-Wolfe signal generated by the largest voids. Potential correlation
between the cosmic microwave background (CMB) temperature variations  with
the quasar voids would confirm the physical nature of voids, and -- in
the future -- could be used to measure amplitude of the large scale
density fluctuations.

The paper is organized as follows. In the next section we present geometric
construction that is used to define voids. In Sec.~\ref{sec:data}  a short
description of the quasar sample used in the investigation is given.
Statistical characteristics of voids found in the data are presented in
Sec.~\ref{sec:voids}. Number of voids, their sizes and shapes are
parametrized by a sphere radius used in the void finding algorithm.
Correlation of the sky position of the largest voids with the CMB
temperature local minima is discussed in Sec.~\ref{sec:vd_cmb}.  Some
peculiarities in void shapes are discussed in Sec.~\ref{sec:vd_shapes}.
Main results are summarized in Sec.~\ref{sec:summary}. Statistics of voids
in a random point distribution and details of the computer algorithm
applied to find  voids are given in the Appendices.


\section{The void -- definition}
 \label{sec:definition}

\begin{figure}
   \includegraphics[width=1.00\linewidth]{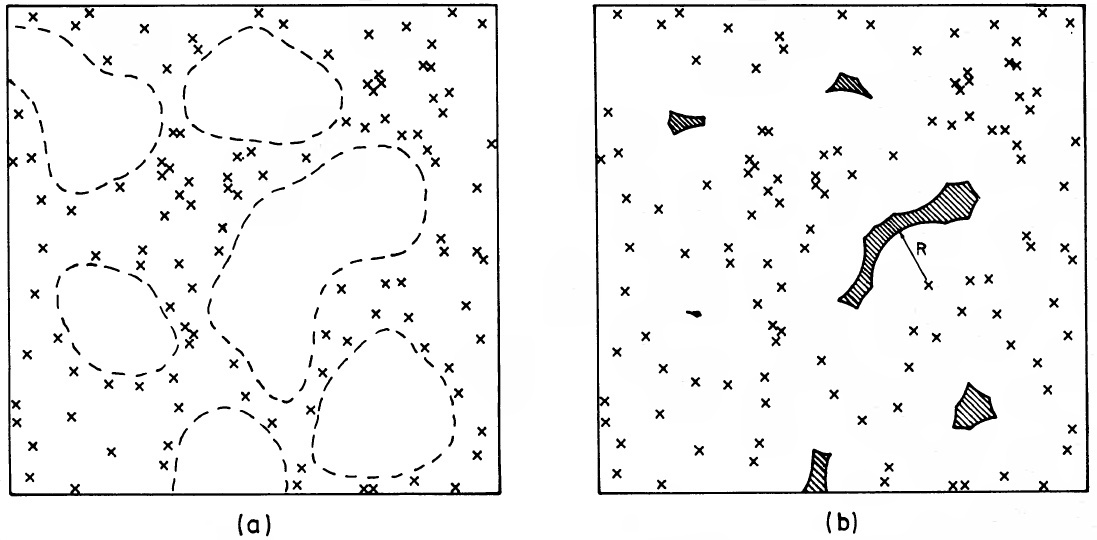}
   \caption{(a) Empty regions isolated by visual inspection and (b) shaded
   areas -- void centres of voids size $R$ \citep[from][]{soltan85}.}
   \label{fig:definition}
\end{figure}

Large and roughly spherical areas devoid of galaxies together with galaxy
clusters, walls and filaments constitute a complex structures of the galaxy
distribution known as a cosmic web. Thus, the notion of cosmic voids was
developed with the advent of the extensive galaxy surveys.  The galaxy
voids are easily identified visually in 3D surveys, but are also revealed
in 2D massive counts such as the Lick galaxy counts.

Space concentration of quasars is much lower than that of galaxies. Thus,
the apparent void in the quasar distribution does not imply presence of the
comparable size galaxy void. Nonetheless, assuming no bias in the large
scale distribution of quasars and galaxies, large quasar voids found in the
present study indicate areas of the lower concentration of galaxies.

To analyze quantitatively even the basic void parameters, such as size and
shape, one should replace the vague concept of void as a rounded empty
volume by more rigorous void definition. Consequently, cosmic voids have
been defined in diverse ways in the past. It has been established that most
galaxy voids are not totally empty, and this observations were incorporated
in various void investigations. In the present study the void is defined as
a region completely devoid of objects.  Figs.~\ref{fig:definition} (a) and
(b), which are taken from \citet{soltan85}, illustrate in 2D a relationship
between: (a) --  the common realization of 'void' notion, and (b) -- the
geometric construction representing the void.  Let us consider the
distribution of $n$ objects in the selected volume $V$. This distribution
contains a void of size $R$ if one can insert into $V$ a circle (sphere in
3D) of radius $R$ with no objects inside.  Shaded areas in
Fig.~\ref{fig:definition} (b) indicate the underlying constructions of
$n(R) = 6$  voids of size $R$.  It is convenient to call each of these
areas a 'void centre'.  Rigorously, the shaded areas are the geometric
places of all the points which are centres for empty circles (spheres in
3D) of radius $R$.

One should note that shape of the void centre contains the information on
the shape of the entire void. Both number of void centres and their shapes
depend on the radius $R$, and --  obviously -- on the statistical
characteristics of the distribution of points. \citet{soltan85} gives the
formula for the expected number of voids (void centres), $n(R)$, as a
function of number of objects, $N$, volume, $V$, and $R$ for the Poissonian
distribution of points in 2D and 3D (see also Appendix~\ref{app:poisson}).


\section{The data}
 \label{sec:data}

\begin{figure}
   \includegraphics[width=1.00\linewidth]{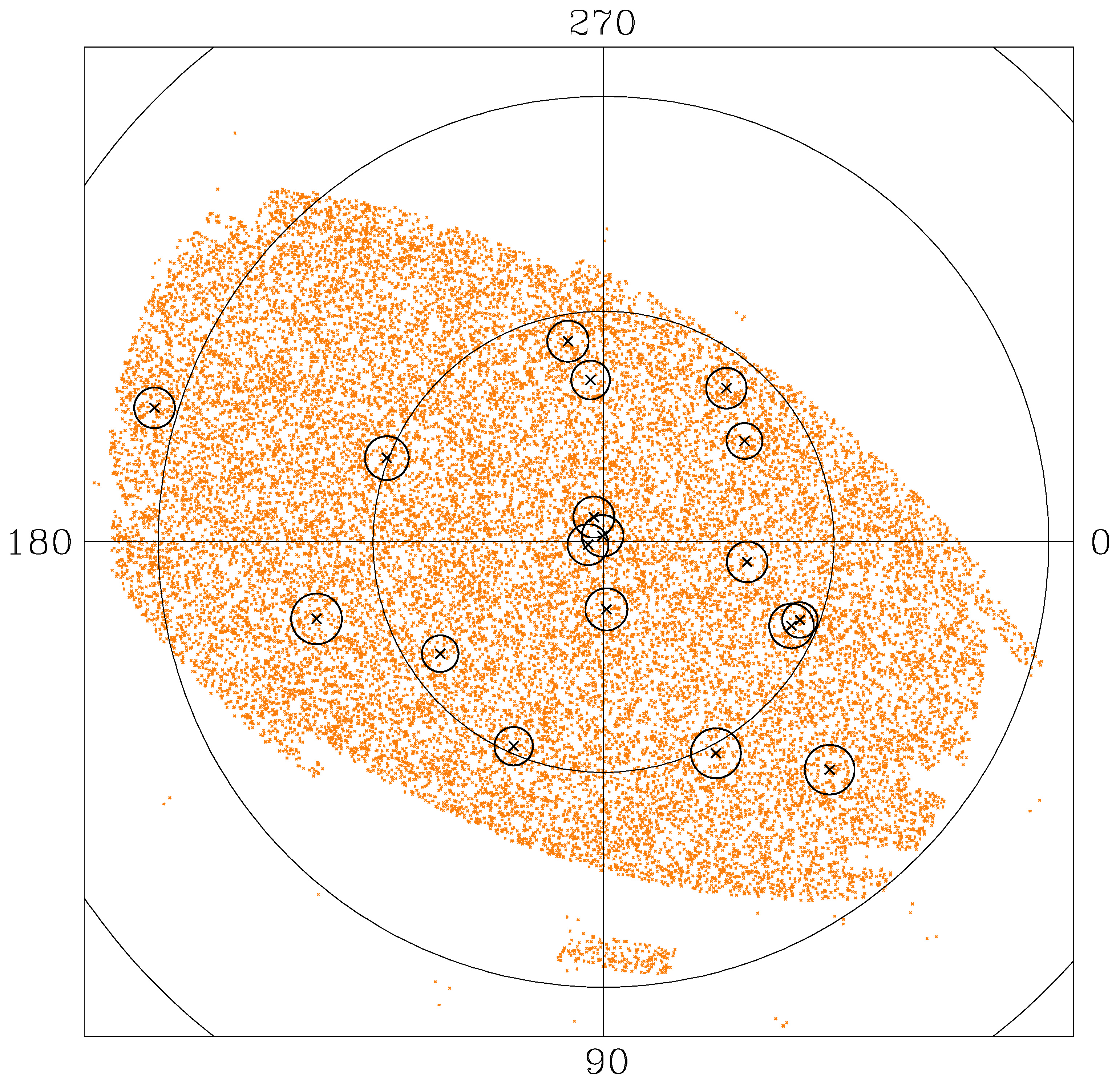}
   \caption{Distribution of $\sim 31000$ quasars from the SDSS DR7 quasar
     catalogue at distances between $2750$ and $4750$\,Mpc in the north
     galactic hemisphere, brighter than $z = 19.50$.
     Galactic latitude circles of $60^\circ$, $30^\circ$ and $0^\circ$
     (arcs) are marked. A sample of voids with radii in the range of
     $177 - 193$ Mpc is shown; circles indicate the
     angular sizes of voids (see Sec.~\ref{sec:vd_cmb}).}
   \label{fig:map}
\end{figure}

The Sloan Digital Sky Survey Quasar Catalogue  is used to study potential
inhomogeneities in the matter distribution on very large scales, i.e.
above $\sim 300$\,Mpc. The fifth edition of this catalogue is presented in
\citet{schneider10}.  The multi step procedure to identify quasars, collect
photometry and spectroscopic redshifts is described in that paper.  Also
the references to papers presenting all the successive releases of the
catalogue are given. Total number of quasars in the catalogue exceeds
$105\,000$, of which more than $90\,000$ lie in the northern galactic
hemisphere. The redshifts range from $0.065$ to $5.46$, and $80$ per cent
of them are between $0.55$ and $2.8$ ($2050$\,Mpc and $6150$\,Mpc).

Generally, the catalogue is magnitude limited what determines the overall
distance distribution. Nevertheless, the distribution of objects in
the redshift magnitude plane is complex what reflects multiple criteria
applied to construct the final list of objects. In particular, quasar
identification and redshift measurement depend on the position of emission
lines in the SDSS photometric system and spectral band pass. Potentially
this effect could introduce spurious variations of the redshift
distribution for the catalogued objects. However the authors stated that
{\it this is not an issue for quasars in this catalogue}. In the paper
we investigate the quasar space distribution in the redshift range
$\sim\!0.8~\div \sim\!1.6$. In this area the resultant selection (see
Fig.~4 of \citealt{schneider10}) apparently generates flat space density
distribution of the catalogued objects.

Void analysis, as other statistics dedicated to the space investigation, is
hindered by the interference of the local effects with the cosmic data.
Although large sections of the catalogue display a high degree of
homogeneity, the surface distribution of objects in several regions of the
celestial sphere is nonuniform and exhibits patches of distinctly different
concentration of quasars. The observational material has been gathered for
several years. Its homogeneity inevitably suffers from instrument-related
biases as well as from the specifics of the data processing. In
particular, selected areas are subject to different magnitude limits. In
effect, most of the distinct features clearly visible in the surface
distribution are not related to the cosmic signal. Our analysis is more
sensitive to various selection effects (in most cases - unrecognized) that
vary across the sky rather than to the radial bias that acts uniformly over
the whole investigated area. Hence, in the present investigation we
introduce an additional magnitude limit of $19.5$ in the $z$ band. Albeit,
this $z$ cut-off decreases the number of quasars in the northern hemisphere
from above $90000$ to $72067$, it effectively reduces conspicuous
surface structures, apparently of the local origin.  The $z$ band, with its
average wavelength of $893$ nm, is used to select a possibly homogeneous
sample because it is least affected by interstellar extinction.
Fig.~\ref{fig:map} shows the distribution of quasars in the north galactic
hemisphere between $2750$ and $4750$\,Mpc selected from the original SDSS
catalogue brighter than $z = 19.50$. Despite its featureless appearance,
the subsequent analysis will show that the space distribution of objects is
not random.

\begin{figure}
   \includegraphics[width=1.00\linewidth]{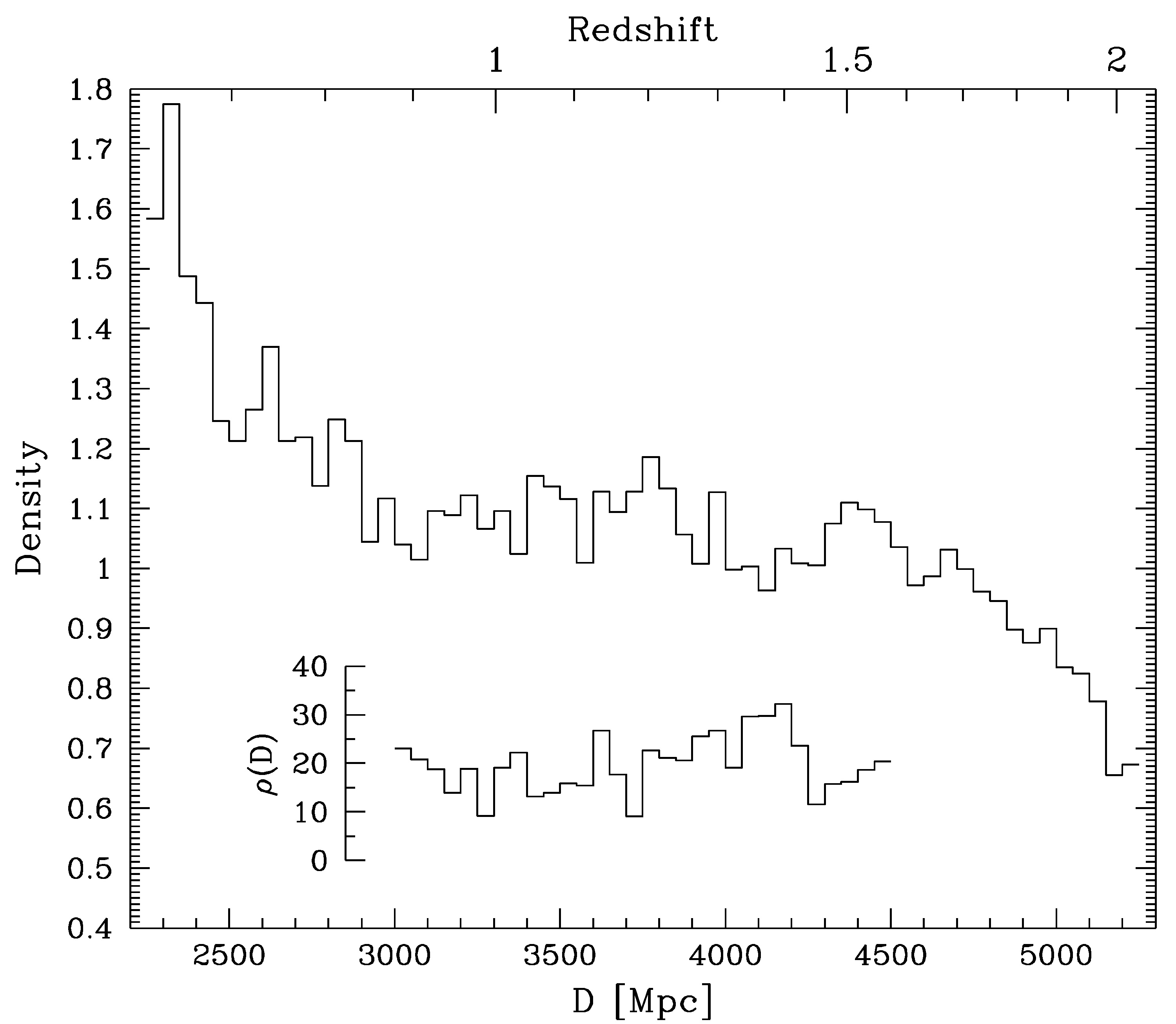}
   \caption{Radial distribution of quasars space density in the north
    galactic hemisphere (arbitrary units). The insert shows the density
    of the void centres in arbitrary units as a function of the distance
    (see Sec.~\ref{sec:voids}).}
   \label{fig:density}
\end{figure}

In Fig.~\ref{fig:density} the space density of the selected quasar sample
is shown as a function of distance. Overall decline of the density with
increasing distance results from an apparent magnitude selection.  A wide
plateau between $3000$ and $4500$\,Mpc comes from  a kind of interplay
between the observational selection and quasar cosmic evolution. This flat
density distribution is helpful for studies of large structures, including
voids.  In the paper we concentrate on this section of the data.  Apparent
depression in the distance range $4000-4300$\, Mpc (centered at redshift of
$1.4$) is discussed below.

\section{Voids in the SDSS quasar distribution}
\label{sec:voids}

\begin{figure*}
   \includegraphics[width=162mm]{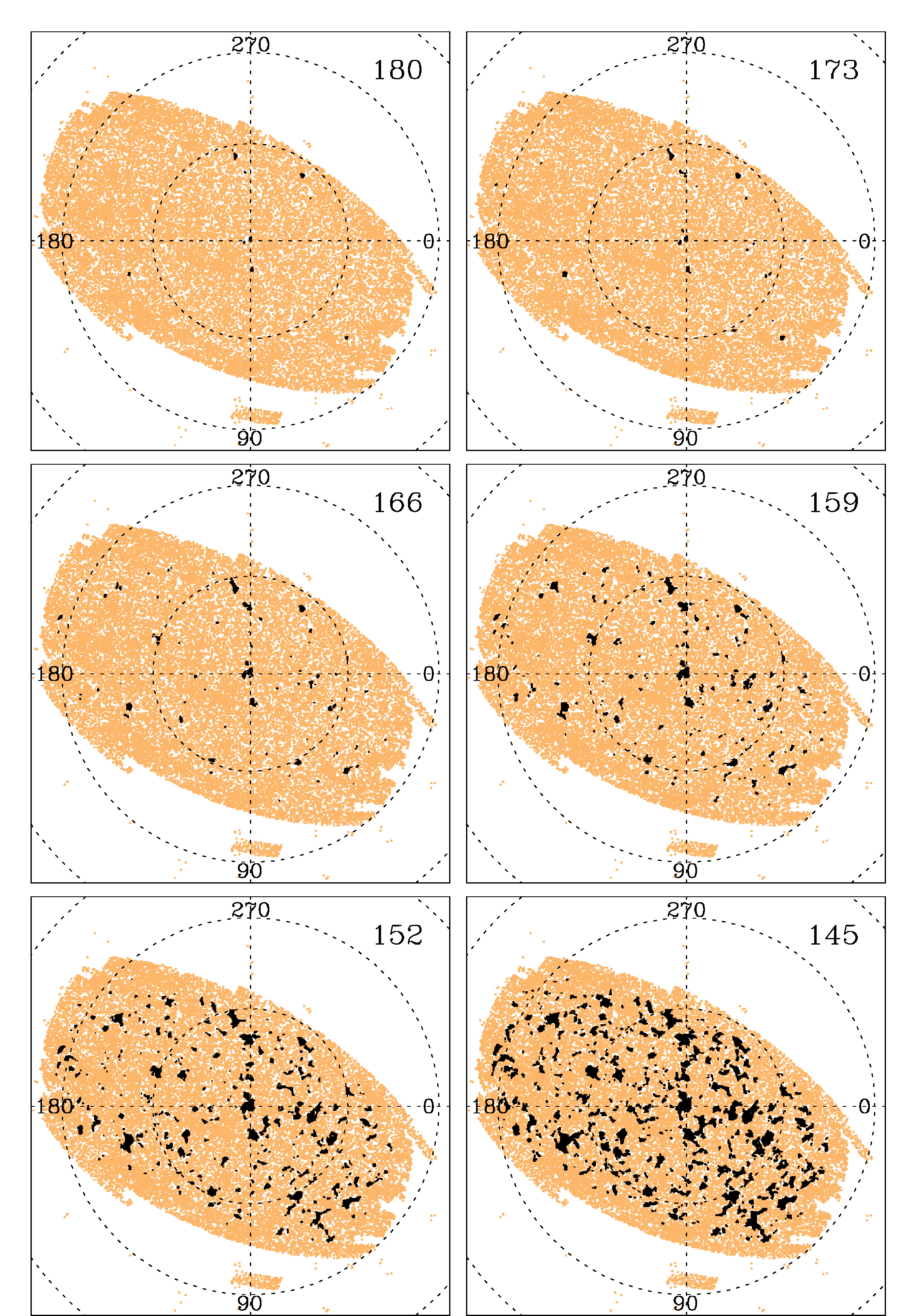}
   \caption{Void centres projected onto celestial sphere for a selection
   of void sizes; numbers in the upper right corners give the void radii
   in Mpc.}
   \label{fig:vd_maps}
\end{figure*}

The void centres in the northern galactic hemisphere were searched for at
distances between $3000$ and $4500$\,Mpc. All the calculations were
performed with the resolution of $1$\,Mpc in 3D. Te radial coordinates were
determined from redshifts assuming the strict Hubble flow in the
$\Lambda$CDM model. The computational details of the void finding algorithm
are described in the Appendix~\ref{app:numerics}. The radius of the largest
void in the investigated area $R = 193$\,Mpc.  Fig.~\ref{fig:vd_maps} shows
the distributions of void centres projected on the sky for the selected
void radii between $180$ and $145$\,Mpc. The number of void centres,
$n(R)$, grows as the radius decreases.  Due to centre mergers the
relationship is non-monotonic.  However, within the radius range covered by
the present computations, a rate at which `new' centres emerge with
decreasing radius,  exceeds the rate of mergers. The effect of void mergers
and the percolation phenomenon is illustrated in Appendix~\ref{app:poisson}
where we show the analytic $n(R)$ relationship over a wide range of radii
for the random distribution of points.


\subsection{Space distribution of voids \label{sec:vd_distr}}

The void centres occupy a small fraction of the investigated volume,
$V = 3.58\times 10^{10}$\, Mpc$^3$, over the entire range of void radii
presented in Fig.~\ref{fig:vd_maps}. At the radius $R = 145$\,Mpc, the
volume of all the void centres amounts to $1.96\times 10^{-3}$ of the $V$.
The present radii are still distinctly greater than the percolation radius
in  the random (Poissonian) distribution of approximately $110$\,Mpc. Thus,
all the selected voids are situated at the 'large void' tail of the void
distribution. In this  radius range, relatively small fluctuations of the
space density of objects result in a strong variations of the void
concentration (see Appendix~\ref{app:poisson}). Consequently, possible
asymmetry of the void distribution on both sides of the galactic longitude
line of $70^\circ - 250^\circ$ (see bottom-right panel of
Fig.~\ref{fig:vd_maps}), if real, may be a result of relatively
small inhomogeneities of the survey rather than the true variations of the
matter density on a Gpc scale.

\begin{figure}
   \includegraphics[width=\columnwidth]{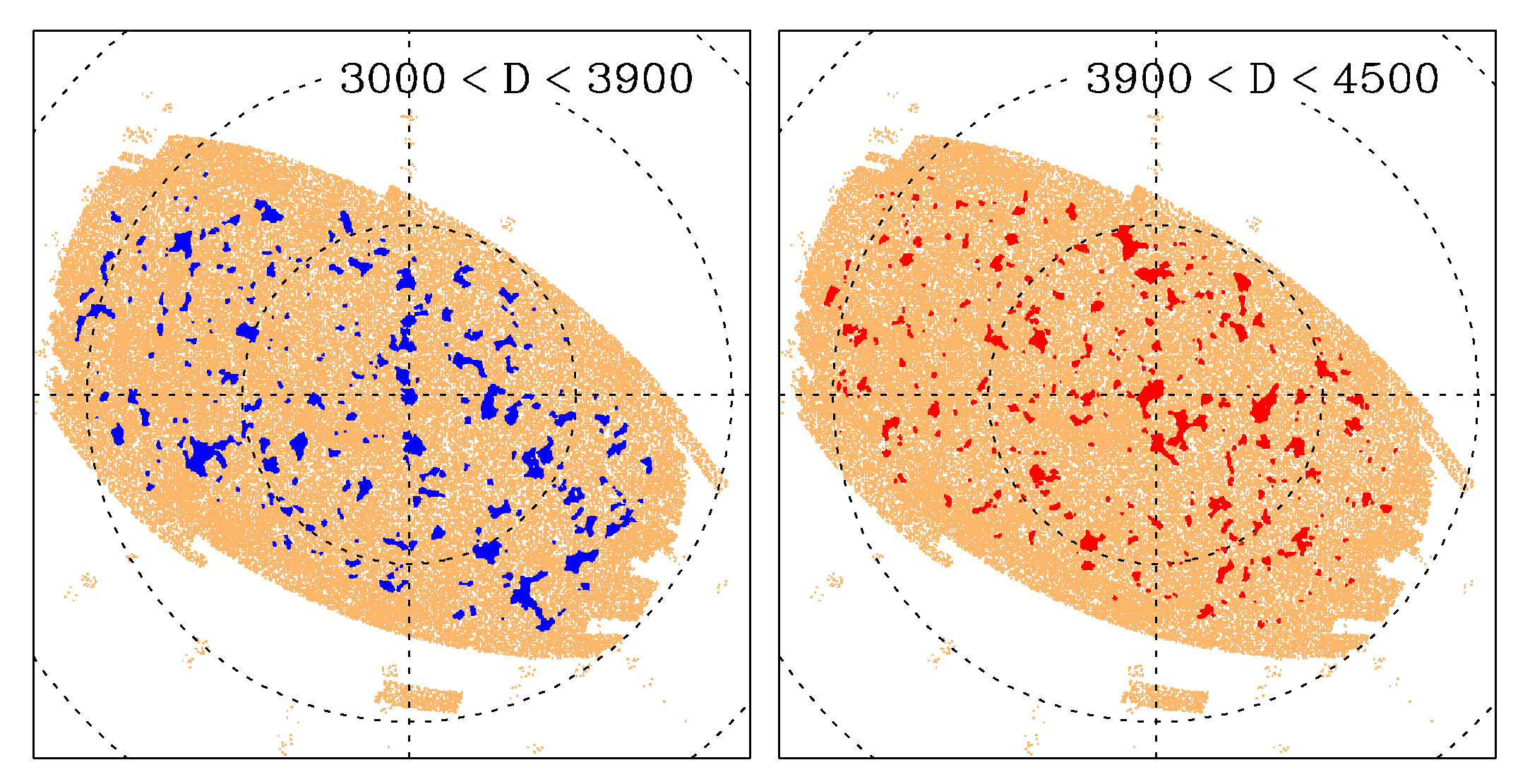}
   \caption{Void distribution at $R=145$\,Mpc in two distance bins
   (in Mpc) indicated in the upper right corners.}
   \label{fig:vd_dist}
\end{figure}

Some insight into this question is provided by exploring the void
distribution along the line of sight. The area between $3000$ and
$4500$\,Mpc is divided into approximately equal volumes. The near field
extends between $3000$ and $3900$\,Mpc, end the far one stretches from
$3900$ to $4500$\,Mpc.  The distribution of void centres found for the
radius of $145$\,Mpc in both regions is shown in Fig.~\ref{fig:vd_dist}.
The number of void centres in the more distant section is larger then in
the near one by more then $26$\,percent. The difference of void
concentration in both fields has been expected because the quasar density
in the more distant field shows a wide minimum centered at the distance of
$\sim\!4100$\,Mpc, or the redshift $z \approx 1.4$
(Fig.~\ref{fig:density}).  Most likely this minimum is caused by the
varying effectiveness of the line identification in quasar spectra
\citep{schneider10}, and is unrelated to the intrinsic quasar distribution.

It appears visually that void centre angular distributions in the celestial
sphere in two distance bins exhibit overall similarities, what would
indicate some inhomogeneities in the SDSS catalogue.  However, it should be
stressed that visually isolated differences of the distribution of void
complexes, are not necessarily significant in the statistical terms.  The
whole population of void centres found at $R = 145$\,Mpc
(Fig.~\ref{fig:vd_maps}, bottom right) does not show clear deviations from
uniformity - the void centres are apparently scattered randomly.  To
investigate this question in a quantitative way, the observed distribution
of separations between the void centres is analyzed. We compare the
distribution of centre pairs in the real and random data sets.

According to the present definition, the void centre is a 3D object, often
of a complex shape. Separation of two void centres is defined as a distance
between their centres of mass.  In the computations, the volume of the void
centre is represented by a set of cubic cells $1$\,Mpc a side (see
Appendix~\ref{app:numerics}).  Due to computer memory constraints, the void
centre is identified and localized in space just by the cells distributed
on the `centre surface'. Only these cells are kept for further analysis.
Because of that, the position of the void centre is described by the centre
of mass of the surface cells.

Two schemes to generate mock catalogues were applied. In the first
one, angular positions of all the centres in the simulated data are the
same as for the real ones, and randomized are only distances to the voids.
We use the bootstrap method, and draw the random distances from the true
distribution. In this case, the randomized void centres preserve some
statistical characteristics of the real data.  In the second method, a
population of void centres found in strictly randomly distributed points is
used.

We examine the distribution of centre pair separations for voids selected
at radius $R=145$\,Mpc. Let $p_{\rm d}(S)$ and $p_{\rm r}(S)$ denote the
numbers of centre pairs with a separation $S$ normalized to the total
number of pairs for the real and randomized void samples, respectively.  A
ratio of both quantities shifted to $0$ for the uncorrelated data
represents a auto-correlation function (ACF) of void centres:

\begin{equation}
\xi(S)= \frac{p_{\rm d}(S)}{p_{\rm r}(S)} - 1\,.      \end{equation}

\begin{figure}
   \includegraphics[width=\columnwidth]{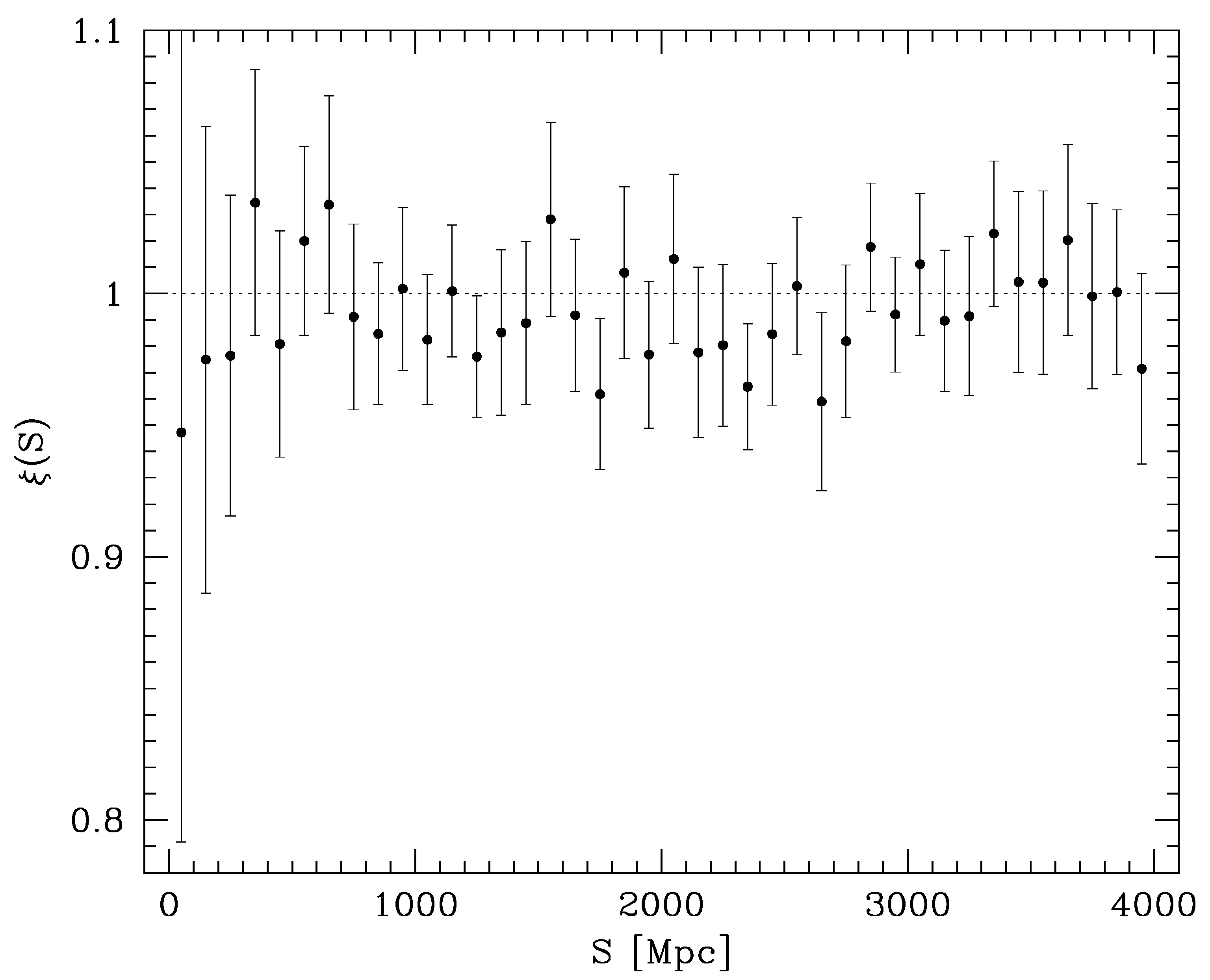}
   \caption{Auto-correlation function of void centres in 3D in $100$\,Mpc
   bins. Error bars show a rms scatter between the mock random data.}
   \label{fig:acf}
\end{figure}

\ni For both randomization schemes no correlation signal was detected.  \ni
Fig.~\ref{fig:acf} shows results for the perfectly random data -- within
statistical fluctuations a ratio $p_{\rm d}(S) / p_{\rm r}(S)$ is equal to
$1$. The pair separations in the random distribution are the average of
$30$ data sets generated using the Monte Carlo scheme. The error bars
represent the rms scatter between the simulations.  No significant
deviations of $p_{\rm d}(S)$ from the random case over a wide range of
separations $S$ indicates that voids are not arranged into larger
structures.  Consequently, the distribution of voids provides no evidences
for the under-dense areas significantly larger than the individual void.

\subsection{Number of voids \label{sec:vd_number}}

\begin{figure}
   \includegraphics[width=\columnwidth]{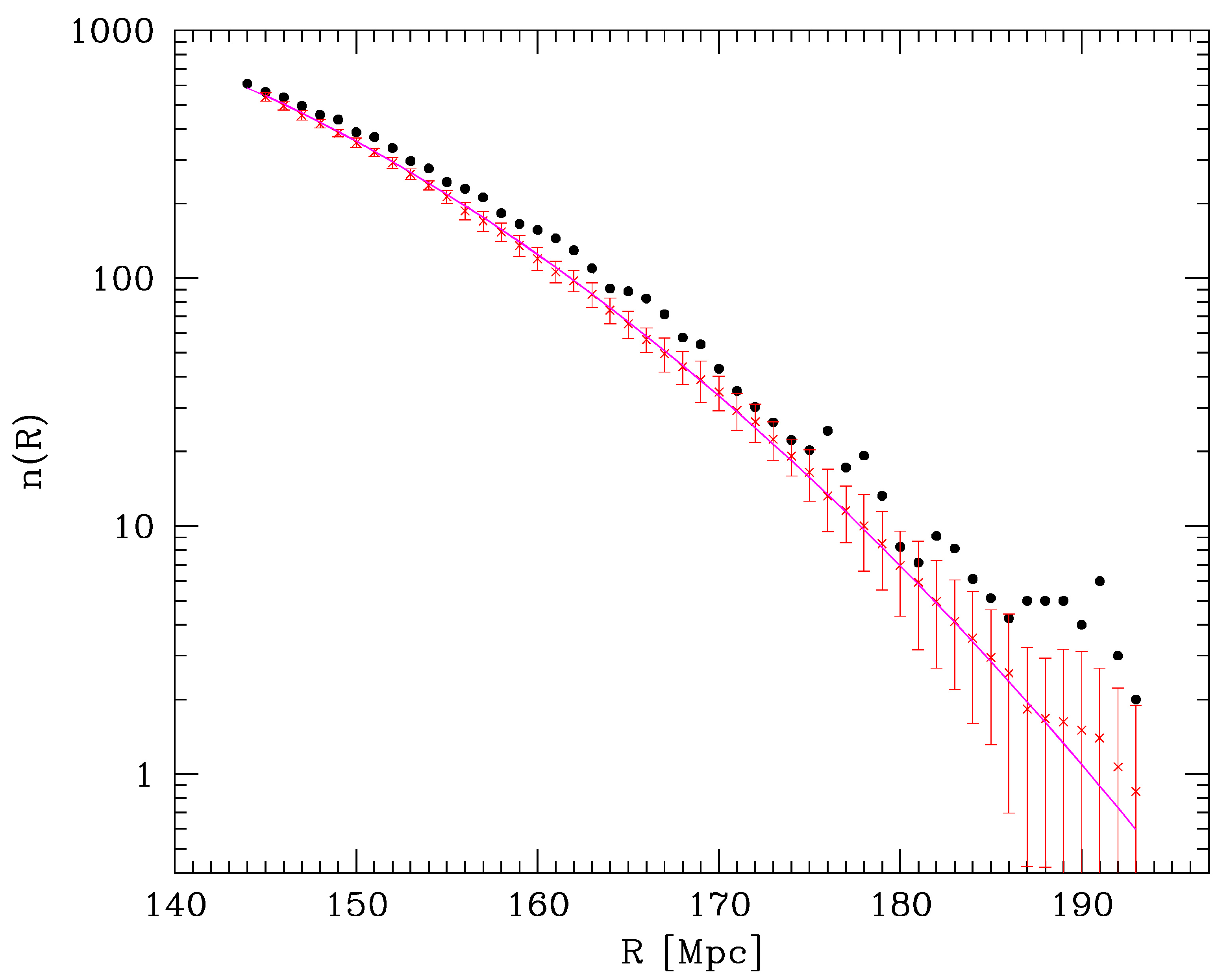}
   \caption{Number of void centres in the SDSS DR7 quasar catalogue: solid
   points are the real data; non-integer values indicate void centres
   reaching the edge of the search area; crosses with error bars --
   average of $30$ mock random distributions (see text for details).}
   \label{fig:vd_number}
\end{figure}

Featureless surface distribution of quasars in the SDSS catalogue
(Fig.~\ref{fig:map}) and close to uniform radial distribution
(Fig.~\ref{fig:density}), indicate that any potential deviations from the
the homogeneous space distribution of quasars are small at large scales.
Therefore, the main objective of this analysis is to assess to what extent
the true quasar distribution in fact differs from the random one. In the
present analysis, the void statistics is used to determine the maximum
linear scale at which the data exhibit non-random characteristics. To
facilitate interpretation of the statistical void parameters, mock
catalogues of uniformly  distributed 'quasars' have been constructed.
Objects in the mock catalogues are randomly distributed in 3D  using pseudo
random number generator within a volume covered by the SDSS. Space density
is equal to the average density in the true catalogue between $2750$ and
$4750$\,Mpc to allow for voids extending beyond the reference distance
range of $3000 - 4500$\,Mpc.  A number of $30$ data sets were created to
gain understanding on statistical scatter of the measured parameters.

The number of separate void centres in the true and simulated data as a
function of radius is shown in Fig.~\ref{fig:vd_number}. In order to
account for the finite survey volume, the void centres are weighted
according to their position relative to the area borders.  The unit weight
is given to all the centres that are completely contained in the analyzed
volume. If the centre touches the boundary of the area at one face, the
weight is reduce to $1/2$, if it touches two faces its weight is assumed to
be equal to $1/4$, and for three faces - $1/8$. Thus, the total number of
voids generally is not an integer. The error bars represent the rms
dispersion determined for $30$ randomized catalogues.  Over the whole range
of void radii the real quasar data accommodate larger number of void
centres, $n_{\rm d}(R)$, than the random simulated distributions, $n_{\rm
r}(R)$. Although, the relative  difference, $[n_{\rm d}(R) - n_{\rm
r}(R)]/n_{\rm r}(R)$, fluctuates due to the stochastic nature of both
distribution, the discrepancy between both quantities seems to decline
steadily at the smaller radii. In the absolute numbers, the excess of true
voids increases with diminishing radius, and stabilizes or begins to
decrease below $r \approx 150$\,Mpc.

The true void excess demonstrates that the void algorithm is an effective
tool to investigate the large scale variations of quasar concentration and
that the catalogued quasars are not distributed randomly in space.
Apparently, the density of quasars is not perfectly constant in the
investigated volume. Because of that, the number of large voids is greater
than the number expected for the random  distribution of objects populating
the same volume with the density equal to the average density of quasars
(see formula \ref{eq:vd_poisson} for the number of voids in the Poissonian
distribution).

It is likely that to some extent residual imperfections of the data
selection procedures introduce some bias in the catalogue which is
responsible for the the $n_{\rm d}(R) - n_{\rm r}(R)$ difference. To
assess, how strongly the cosmic signal contributes to this difference,
other statistical properties of voids are investigated in the following
sections.


\section{Voids vs. CMB}
\label{sec:vd_cmb}

Fluctuations of matter distribution on very large scales influence the
cosmic microwave background (CMB) what is known as the Integrated
Sachs-Wolfe effect (ISW). CMB photons crossing mass agglomerations, as well
as broad depressions of the mass density. gain or lose energy.  Amplitude
of the effect is defined by the net change of height of the potential hill
(or depth of the well) during the photon travel time. In the matter
dominated flat universe the time evolution of density fluctuations in the
linear approximation is balanced  by the matter dilution caused by the
Hubble expansion and the effect is cancelled. In models with the
cosmological constant the expansion of the universe is not matched by the
evolution rate of density fluctuations and the ISW effect turns up. 

From the observation point of view the ISW effect is still debatable.
Although, the large structures (both superclusters and supervoids) found in
extensive galaxy surveys, such as SDSS DR6 \citep{adelman08}, seem to
correlate with the CMB temperature variations, the reported amplitude of
the signal (e.g.  \citealt{granett08,ilic13,kovacs17}) is substantially
larger than that expected for the ISW effect (e.g.
\citealt{hernandez13,hotchkiss15}).

Here, we explore the potential relationship between the voids found in the
quasar catalogue and the CMB temperature fluctuations under the assumption
that the correlation of both distributions is generated via the ISW effect.
However, to confirm the existence of the under-dense regions associated with
our voids, the nature of this correlation is not crucial.  Obviously, if
the observed void excess results from the catalogue deficiencies generated
locally, one should expect no correlations of the detected voids with the
CMB temperature. Moreover, the void excess can be generated by various
kinds of fluctuations of the matter density that also do not introduce void
- CMB correlation. For example, if the void excess is produced by
  variations of the density field on scales much larger than the present
voids, the number of detected voids will exceed that for the Poissonian
field, but a topology of individual voids will be defined just by
geometrical structures unrelated to the local matter density. Only the
physical connection of some voids with the true matter density depressions
generates the sought correlation.

Figure~\ref{fig:vd_number} shows that the number of voids detected in the
quasar distribution is greater than that expected in the perfect Poissonian
case, but the relative difference of both quantities is rather modest.
Thus, the majority of voids is not associated with the areas of lower than
average matter density. Nevertheless, the systematic excess of the void
number over a wide range of radii could indicate that some fraction of
large voids is genetically related to the under-dense regions. We
investigate in detail a possible coincidence of `cold spots' in the CMB
temperature map with the population of voids found in the present
investigation. We note the obvious fact that the number of voids selected
at two void radii are correlated.  This is illustrated in
Fig.~\ref{fig:vd_maps}, where all the voids selected at one radius show up
also in the void  maps constructed at all smaller radii. To obtain
independent sets of void centres for different radii, we apply the
following procedure. First, a void of the largest radius is found.  It is
centred at galactic coordinates $(l, b) = (216^\circ, 89^\circ)$ at the
distance of $4139$\,Mpc. Then the centre of the next in size void is
searched for, excluding the volume occupied by the first one.  The minimum
void centre separation equal to the radius of the larger void is introduced
to eliminate the situation where several void centres are clustered in the
small area. Such tight void group in fact represents potentially a single
under-dense region. The second void with radius of $192$\,Mpc is centred at
$(l, b) = (309^\circ, 64^\circ)$ at the distance of $4234$\,Mpc. The
procedure was repeated for all the voids down to $r =145$\,Mpc, the
smallest voids investigated here, and provided a list of $568$ void
centres.  A sky distribution of a sample of $18$ largest voids selected in
this way in the radius range $193 - 176$\,Mpc is shown in
Fig.~\ref{fig:map}.

\begin{figure}
   \includegraphics[width=\columnwidth]{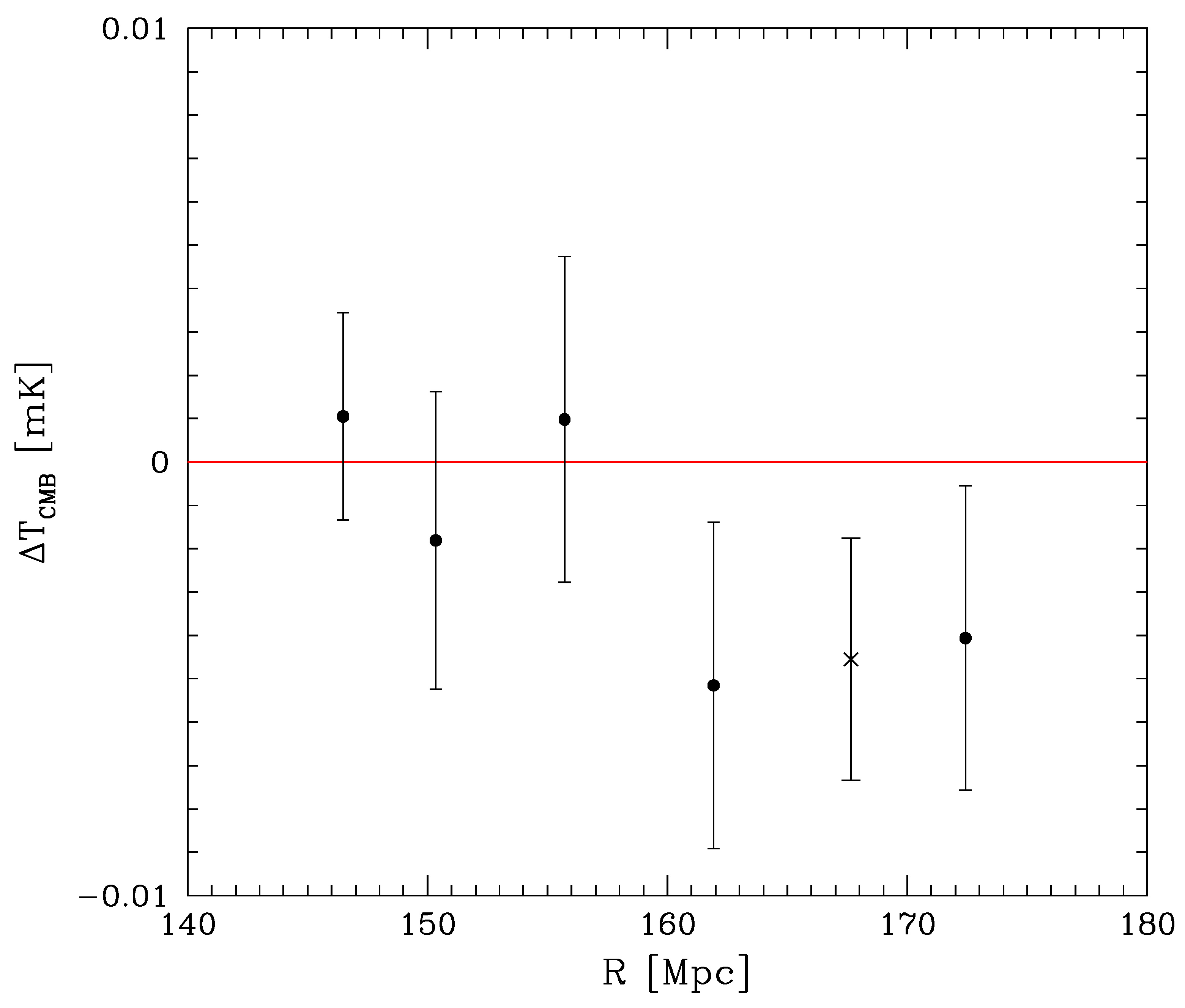}
   \caption{The CMB temperatures of all 
   the voids grouped according to the void radius. The
   ordinate $\Delta T_{\rm CMB}$ shows the difference between the
   temperature in the direction of the void and the temperature in the
   annulus surrounding the void (see the text for details).}
   \label{fig:vd_cmb_all}
\end{figure}

\begin{figure}
   \includegraphics[width=\columnwidth]{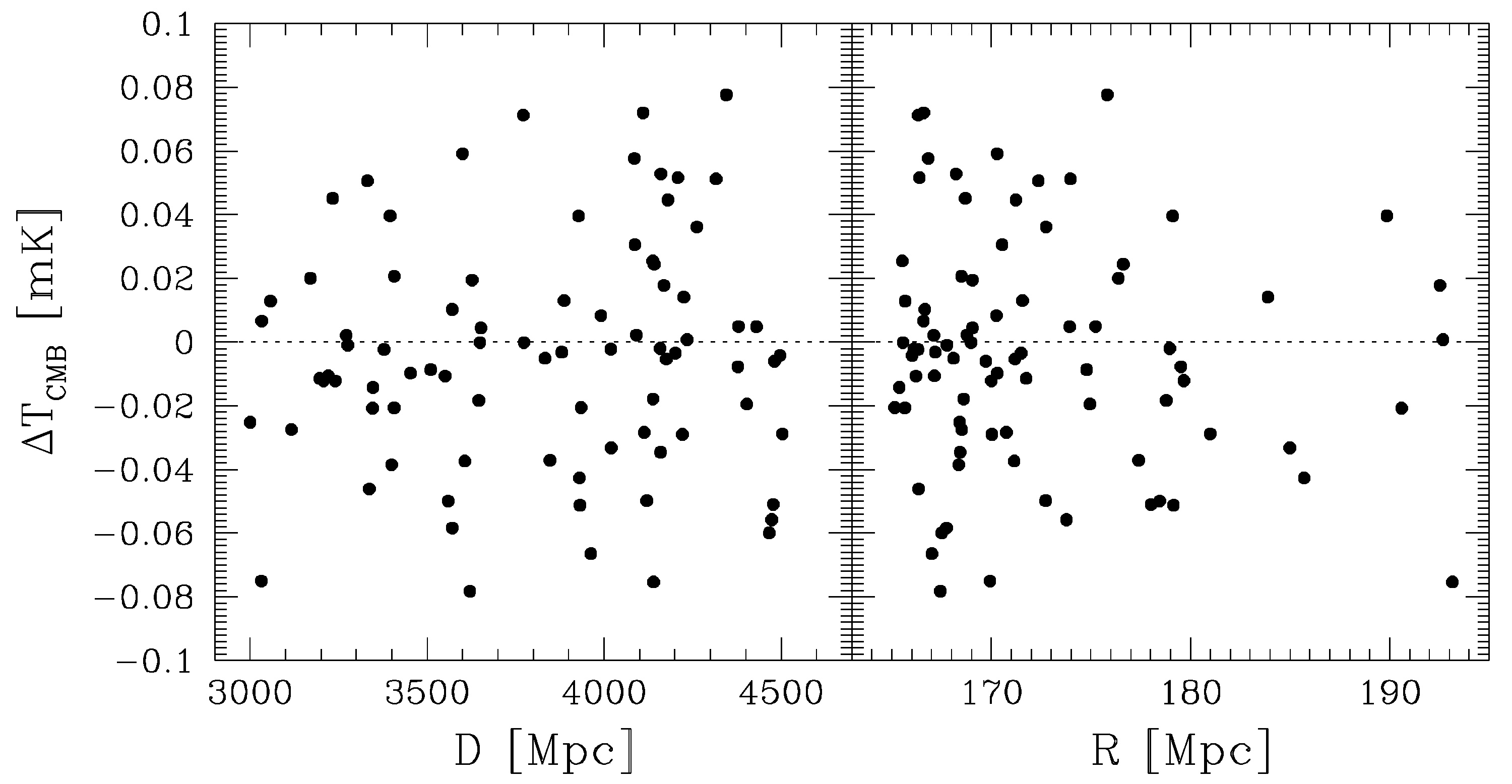}
   \caption{Distribution of void temperatures vs. distance and radius
    of voids with $R \ge 165$\,Mpc.}
   \label{fig:vd_cmb_165}
\end{figure}

The number of the observed voids with smaller radii approaches that
predicted for the Poissonian case. These voids result from the purely
random quasar structures and do not produce the ISW effect. Thus, the
expected ISW signal averaged over the whole void population is highly
diluted.  Moreover, the fluctuations of the CMB temperature in the WMAP
data at all the interesting scales are much larger than the expected ISW
amplitude created by voids, what strongly reduces the signal-to-noise (S/N)
ratio of the present ISW signal detection.

The CMB temperatures at direction of voids are calculated using the $9$
year WMAP CMB maps available at
LAMBDA\footnote{\verb|https://lambda.gsfc.nasa.gov/product/map/dr5/ilc_map_get.cfm|}
which is  a part of NASA's High Energy Astrophysics Science Archive
Research Center. We used  WMAP's standard Res 9 HEALPix projection with
$\sim 7$ arcmin resolution.  For each void the CMB temperature was
determined. The WMAP data were averaged over a circular sky area positioned
at the void centre. One can expect that the highest amplitude of the ISW
signal coincides with with the void centre and decays outside.  To maximize
the S/N ratio, the radius of the area was delineated as $\theta_{\rm CMB} =
\varkappa\, \theta_{\rm void}$, with $\varkappa = 0.9$, where $\theta_{\rm
void}$ is the angular radius of a void (different values of $\varkappa$
give either the weaker signal or the stronger noise).  The amplitude of the
ISW signal was defined as the difference between the void temperature and
the average CMB temperature in the surrounding area.  The CMB map was
smoothed using a spherical harmonic filter of degree $l = 9$.  The low
value of the degree $l$ was chosen to produce sufficiently flat temperature
map around each void. The reference temperature to calculate the ISW signal
was assumed as the average temperature of the filtered CMB data in the
annulus surrounding the void. The inner and outer annulus radii of
$\theta_{\rm CMB}$ and $3\,\theta_{\rm CMB}$ were taken.

The $\Delta T_{\rm CMB}$ plotted in Fig.~\ref{fig:vd_cmb_all} show the
differences between the void and reference temperatures in the set of $568$
voids defined above. The data are divided into $5$ bins according to the
void radius:$145 \le R < 148$\,Mpc, $148 \le R < 153$\,Mpc, $153 \le R <
159$\,Mpc, $159\le R < 165$\,Mpc, and $R \ge 165$\,Mpc.  The  bins contain
respectively: $125$, $155$, $136$, $69$, and $83$ voids. The cross shows
the temperature difference for the merged two bins with largest radii ($R
\ge 159$\,Mpc). The $1\,\sigma$ uncertainties are estimated using the
simulations. The void and reference temperatures were obtained for a set of
$30$ randomized catalogues. The data have been processed in the same way as
the original material, and the rms scatter of the corresponding $\Delta
T_{\rm CMB}$ distributions in the mock catalogues is shown as the error
bars.  Details of the radial distribution and radii of voids with $R \ge
165$\,Mpc are displayed in Fig.~\ref{fig:vd_cmb_165}.

The distribution of temperatures in large voids is not symmetric with
respect to the $\Delta T = 0$ line.  The average negative $\Delta T_{\rm
CMB}$ signal for voids with $R \ge 159$ differs from zero by more than
$1.64\,\sigma$, and in the highest bin of $R \ge 165$\,Mpc the net
temperature in $50$ voids is negative.  Using the binomial distribution and
assuming equal probabilities of temperatures below and above the average, a
chance that at least $50$ of $83$ temperatures drawn at random will be
negative amounts to $0.039$. In statistical terms, significance of this
asymmetry is not very high.  Nevertheless, the data are consistent with the
correlation between the void distribution and the CMB temperature
depressions, and it is legitimate to assume that some voids indeed coincide
with the areas of lower than average matter density.  Unfortunately, large
intrinsic scatter of the CMB temperature maps strongly impedes assessing
the amplitude of the ISW signal.  This in turn, prevents us from imposing
restrictive constraints on the density distribution in voids.
Consequently, our objective in the subsequent calculations, is to examine
to what extent the present estimates of $\Delta T_{\rm CMB}$ are consistent
with the existing data on the large scale matter distribution derived from
N-body simulations.

The relationship between the cosmic structures and the CMB temperature
variations generated by the ISW effect has been broadly discussed in the
past.  In particular, a measurable correlation of the void distribution
with depressions of the CMB temperature is expected in models with
non-vanishing cosmological constant \citep[e.g.][]{nadathur16}, although
the predicted ISW signal is typically much weaker than those reported in
the literature, \citep[see][]{nadathur14}.  Amplitude of $\Delta T$
depletion produced by the void depends on the time evolution of the
gravitational potential, $\Phi(r)$, along the CMB photon path.

The distribution of $\Phi(r)$ in the vicinity of voids was investigated by
\citet{nadathur17} using the N-body simulations. They found that for large
voids, $\Phi(r)$ scales in a simple way with the void radius and the
average galaxy density contrast within a void, $\delta_g =\rho_{\rm
gv}/\rho_{\rm gl}-1$, where $\rho_{\rm gl}$ and $\rho_{\rm gv}$ are the
global and void galaxy number densities, respectively. The approximate
galaxy bias factor in voids is estimated at $2$.  In the following we
assume that these scaling relationships apply to the present voids, albeit
the \citet{nadathur17} investigation concentrates on lower redshifts and
smaller void sizes than those considered in the present paper.  Thus, the
subsequent calculations have only indicative character.

To assess capabilities to measure the large scale inhomogeneities of the
matter distribution using the present void algorithm we apply the linear
formula for the ISW temperature change \citep{nadathur16}:

\begin{equation}
\frac{\Delta T}{T} =  -2 \int_0^{z_{\rm LS}} a(z) [1 - f(z)]\,\Phi\,dz\,,
\label{eq:isw}
\end{equation}

\ni where $z_{\rm LS}$ is the redshift of the last scattering, $a(z)$ -- 
the cosmic scale factor, and 

\begin{equation}
f = \frac{d \ln D}{d \ln a}\,,
\end{equation}

\ni is the density fluctuation growth rate, where $D(a)$ is the linear
evolution of the matter density perturbation.

One can expect that the highest  $\Delta T_{\rm CMB}$ amplitudes are
generated preferentially by the largest voids. In the present
investigation, however, the estimates of the average temperature signal are
strongly affected by sampling errors. Because of that we include in the
calculations all the voids with $R \ge 159$\,Mpc.  The average void radius
in a sample of all the voids with $R \ge 159$\,Mpc is $\overline R =
167.7$\,Mpc, and the average void distance $\overline d = 3820$\,Mpc
($\overline z \approx 1.2$). Using \citet{nadathur17} Eqs (6) and (10) with
scaling coefficients in their Table B2, we get the distribution of
potential, $\Phi(r)$, associated with the our 'average' void for different
amplitudes of the galaxy contrast $\delta_g$. Then, the integral in
Eq.~\ref{eq:isw} is calculated in the $\Lambda$CDM cosmological model with
parameters specified in the Introduction. The formulae for the linear
growth rate, $D(a)$ were taken from \citet[][p.\,49--51]{peebles80}.

To compare the model with our $\Delta T_{\rm CMB}$ estimates, we note that
only a fraction of the voids is actually associated with the low density
areas, while majority of voids results from random quasar configurations.
Therefore, we define the average temperature signal produced by a true
fluctuation of the matter distribution as:

\begin{equation}
\Delta T_{\rm u} = \eta\,\frac{\Delta T_{\rm v} \cdot n_{\rm v}}
                              {n_{\rm v} - n_{\rm rv}}\,,
\label{eq:dilution}
\end{equation}

\ni where $\Delta T_{\rm v}$ and $n_{\rm v}$ are the average temperature
deficit in a void and the number of voids in the sample, and $n_{\rm rv}$
denotes the number of voids expected for the random distribution. The
coefficient $\eta$ takes into account partial overlapping of voids, and is
equal to the ratio of the total solid angle covered by voids to the summed
up area of all the voids. For $R \ge 159$  and $\varkappa = 0.9$ we have
$\eta = 0.753$, $\Delta T_{\rm v} = -0.00465 \pm 0,0028$, and $n_{\rm v} =
152$. The number of voids expected in the random distribution $n_{\rm rv} =
126.9 \pm 10.2$ is derived from $30$ mock catalogues. Thus, the temperature
depletion associated with the under-dense regions $\Delta T_{\rm u} = -0.022
\pm 0.016$\,mK. According to Eq.~\ref{eq:isw} the temperature depletion at
the level of $-0.022$\,mK is produced by a void with the average density
contrast of $\sim -0.62$. Although, the data on the galaxy distribution at
redshifts under consideration are scarce, such high density contrasts seem
unlikely. Also, the galaxy bias factor generally grows with
redshift (e.g. \citealt{papageorgiou12}) and the corresponding galaxy
concentration contrast would be even higher.
Assuming large (as compared to the
\citet{nadathur17} assessments) matter density contrast of $-0.3$,
the temperature depletion according to Eqs~\ref{eq:isw} and
\ref{eq:dilution} from our `average' void
$\Delta T_{\rm u} = -0.006$\,mK, and 
$\Delta T_{\rm v} = 0.0013$\,mK.
In view of the large CMB temperature variance,
the `dilution' of true underdense regions among the random quasar voids 
drastically reduces ability of the
present method to investigate the ISW effect from voids.

We stress, that all these estimates are based on the extrapolation of the
\citet{nadathur17} scaling relations for the relevant void parameters.
Their distribution of void radii peaks at $\sim 40$\,Mpc with no voids  of
$R$ above $90$\,Mpc. \citet{nadathur12} assess that in $\Lambda$CDM models
$\sim 100$\,Mpc structures at $z \approx 0.5$ generate the ISW signal of
$\lesssim 0.002$\,mK. The hypothetical structures reported here exceed
$300$\,Mpc, and proportional higher temperature effect is expected.
Such structures are, however, extremely rare. In the volume of $\sim
3.6\cdot 10^{10}$\,Mpc, the number of under-dense regions associated with
voids is estimated at $27\pm 10$. Thus, there is less than one such object
in a cube of side $1100$\,Mpc.  It underlines the potential role of large
void searches in the quasar distributions for the investigation of matter
density fluctuations at scales much larger than accessible in the galaxy
catalogues.


\section{Void shapes} \label{sec:vd_shapes}

\begin{figure*}
   \includegraphics[width=185mm]{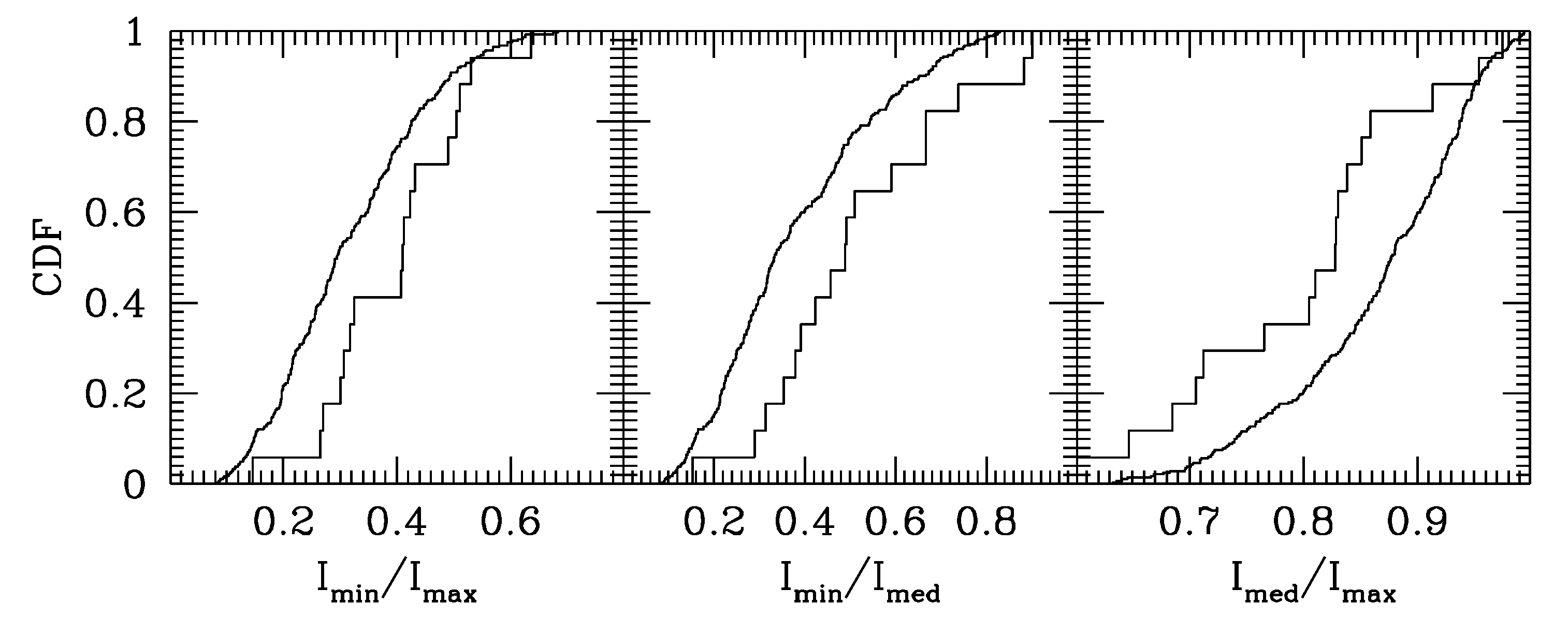}
   \caption{Cumulative distribution functions (CDF) of ratios
   of the principal moments of inertia:
   minimum to maximum, $I_{\rm min}/I_{\rm max}$,
   minimum to medium. $I_{\rm min}/I_{\rm med}$, and
   medium to maximum, $I_{\rm med}/I_{\rm min}$
   for $17$ largest voids in the real data and for $242$
   voids in $30$ sets of the randomized data.
   The voids were selected at the void radius $R=145$\,Mpc.}
   \label{fig:inertia}
\end{figure*}

A shape of the void centre is defined by cells distributed in the centre
surface, as described in the Sec.~\ref{sec:vd_distr}.  Obviously, shapes
evolve with the void radius $R$. One can expect that statistical
investigation of void centre structures should give also some insight into
the typical properties of voids understood as volumes of space free of
quasars. A moment of inertia is a natural tool to study the basic geometric
properties of 3D structures. For that purpose, the void centre is
considered as a rigid body.

Since the void centre is here represented only by cells in the outer layer
of the spatially extended centre structure, the moment of inertia pertains
just the `mass distribution' of the surface of the centre.  Principal
moments of inertia and principal axes (eigenvalues and eigenvectors of the
moment of inertia tensor) of void centres were calculated for the real and
randomized data. Gross features of the centres, such as sizes, ratios of
the principal moments, and directions of principal axes in both data sets
were investigated. No conspicuous distinctions between the sets have been
found, albeit some marginally significant differences in the centres shapes
are present.

The distribution of quasars is a Poissonian stochastic process, and voids
in the real data not necessarily reflect statistically significant large
scale depression of the matter distribution. One can expect that preferably
only the largest voids that are found occasionally in the real catalogue
indicate areas of lower matter density. Among voids selected at
$R=145$\,Mpc, $17$ have `surface' (as defined above) containing more that
$1.5\cdot 10^5$ cubic $1$\,Mpc cells.  In the $30$ sets of the randomized
distributions, only $242$ such large voids have been found.
Fig.~\ref{fig:inertia} shows the cumulative distribution functions, CDF,
of the principal moments of inertia ratios for the real data ($17$ step
functions) and the randomized data. 

\begin{figure}
   \includegraphics[width=\columnwidth]{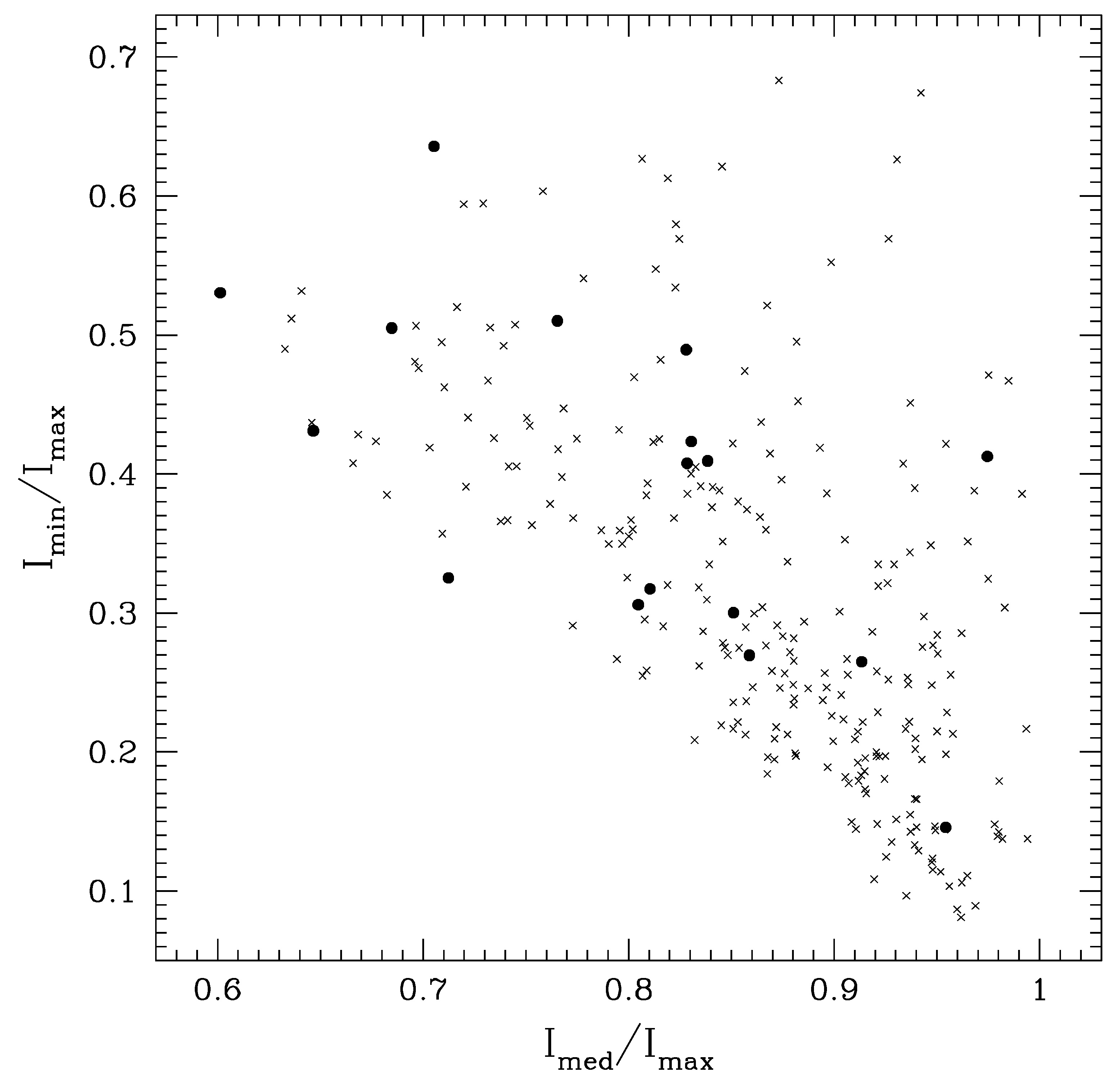}
   \caption{Distribution of principal moments of inertia ratios
   for void centres containing more that $1.5\cdot 10^5$ cells
   (see text for explanation) in the real data (full dots) and
   in $30$ randomized catalogues (crosses).}
   \label{fig:inrt_2D}
\end{figure}

A two sample K-S test applied to the distributions of the principal moments
of inertia, $I_{\rm max}$, $I_{\rm med}$, $I_{\rm min}$, reveals possible
differences of void centre shapes in the data and in the simulations. In
Fig~\ref{fig:inertia} the cumulative distribution functions (CDF) of
$I_{\rm min}/I_{\rm max}$, $I_{\rm min}/I_{\rm med}$, and $I_{\rm
med}/I_{\rm max}$ for the true and random voids are shown. The
distributions differ at a significance level of $0.039$, $0.032$ and
$0.006$, respectively. Relationships between three principal moments are
also visualized in Fig.~\ref{fig:inrt_2D}, where the distribution of
$I_{\rm min}/I_{\rm max}$ vs. $I_{\rm min}/I_{\rm med}$ is shown for the
void centres in the quasar data and in the mock catalogues. To compare both
distributions we apply two-dimensional version of the K-S test developed by
\citet{peacock83}. The test indicates that real and random data differ at
significance level of $0.0073$. The signal in the Peacock's test is
produced by the apparent excess of void centres in the top-left quadrant of
Fig.~\ref{fig:inrt_2D}. Judging from  this feature, it implies that the
shapes of centres are generally less elongated (or more spherical) as
compared to the random case. One should keep in mind, however, that the
actual shapes of void centres of the large voids are highly different from
the `regular' shapes of triaxial ellipsoids.

\section{Concluding remarks}
\label{sec:summary}

We study the space distribution of quasars in the SDSS to assess the matter
distribution on very large scales. A suitable quasar void finding algorithm
allowed for the investigation of void sizes and shapes. An existence of
structures involving groups of voids were also examined. The SDSS quasar
catalogue, taken as a whole, is subject to various selection biases. A
section of the catalogue that covers the redshift range of roughly $0.8 -
1.6$ was used in the paper. Although, this area seems the best fitted for
this kind of investigation, the data also suffered from imperfections, what
limited the present study. Potential inhomogeneities of the SDSS could
affected overall space distribution of voids.  Because of that we
concentrate on void characteristics that are fairly immune to the survey
deficiencies.

It is shown that the distribution of void sizes is inconsistent with the
random distribution of quasars. The excess of the number of voids is
observed for  diameters above $\sim\!300$\,Mpc.  We examine the largest
voids, since they most likely coincide with the underlying large scale low
matter density (baryonic and dark) areas. To relate the present voids with
the true depressions of the matter density, we investigate the angular
correlation between the voids and the CMB temperature distribution. Such
correlation is expected due to the ISW effect. It is found that the average
temperature in the direction of large voids is lower then in the
surrounding areas by a few $\mu$K in rough agreement with the ISW mechanism
in the $\Lambda$CDM model. However, the statistical significance of
the detection is too low to perform a quantitative analysis of the ISW
effect. Excess of voids with the negative temperature deviation over those
with the positive one among $152$ voids with $R \ge 159$\,Mpc is
significant at $0.039$ level. We conclude that the low significance of both
statistics results from the fact that extraneous variations of the CMB
temperature are much larger that the measured amplitude of the ISW signal.

The space autocorrelation function of void centres is determined.  The ACF
amplitude is consistent with no correlation signal over a whole accessible
range of void separations. However, this conclusion is not highly
restrictive due to large uncertainties of the ACF estimate. To improve the
voids statistics, data covering wider redshift range would be required.

In statistical terms, the shapes of the quasar void centres define space
structures of under-dense areas. Thus, investigation of quasar voids could
provide valuable information on the large scale matter distribution. The
observed void centre shapes and those found in the random distributions are
statistically different, although the differences are not high. Shapes
here are defined solely by the ratios of principal moments of inertia
$I_{\rm min} / I_{\rm max}$ and $I_{\rm med} / I_{\rm max}$.  Assuming that
distributions of these parameters describe a population of triaxial
ellipsoids, real objects tend to be more spherical as compared to the
simulated ones.  However, the true shapes of the largest void centres in
Fig.~\ref{fig:vd_maps} (bottom right panel) strongly differ from
ellipsoids, and conclusions based on such approximation should be treated
cautiously. We plan to extend the investigation of voids using quasars in
other redshift ranges. Broader observational basis should help to clarify
also this point of the present paper.


\section*{acknowledgements}
I thank the anonymous reviewer for valuable recommendations
that greatly helped me to improve the material content of the paper.

\bibliography{soltan_sh}

\appendix

\section[]{Voids in Poissonian distribution}
\label{app:poisson}

In Appendices we use terms `void' and `void centre' as defined in
Sec.~\ref{sec:definition}.

Although the space distribution of quasars listed in the SDSS catalogue is
different from the random (Poissonian) one, it is instructive to compare
the void centres characteristics of the real data with this idealized case.
Additionally, analytic formula for the number of voids in the Poissonian
case derived in \citet{soltan85} allows us to check quality of the void
finder computer algorithm.

\begin{figure}
   \includegraphics[width=\columnwidth]{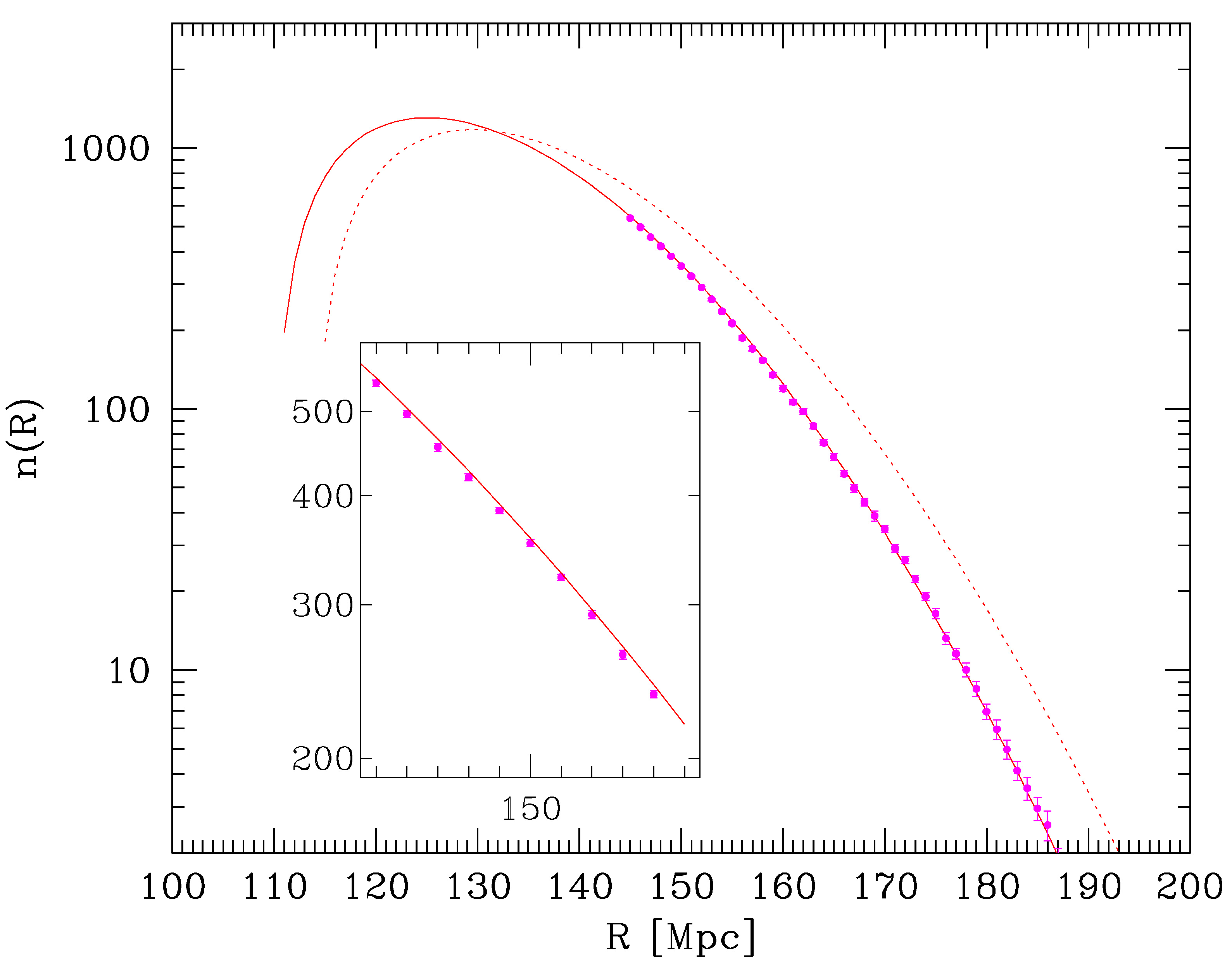}
   \caption{Number of voids for the random distribution of points in
    a volume  $V = 3.58\cdot10^{10}$\,Mpc$^3$ with the concentration
    $\lambda = 5.15\cdot10^{-7}$\,Mpc$^{-3}$: solid curve -- the expected
    number according to formula~\ref{eq:vd_poisson}, points with the
    error bars -- computed number using the present algorithm from the
    mock data generated by the Monte Carlo method; the insert shows
    the data for $145 \le R \le 155$\,Mpc in greater details; dotted
    curve -- the expected number of voids for the number density $10$
     percent lower, i.e. $\lambda = 4.64\cdot10^{-7}$\,Mpc$^{-3}$.}
   \label{fig:sim_vd_number}
\end{figure}

A number of random mock catalogues in the SDSS area were generated using the
Monte Carlo scheme. Then, the void centres were found applying the same
method as for the real data (see below).  Results averaged over $30$ mock
catalogues are shown in Fig.~\ref{fig:sim_vd_number} with dots. The error
bars represent the rms dispersion of $30$ data sets around the average number
of centres $n(R)$, divided by the square root of the number of sets.  The
volume searched for void centres, $V$, and the space density of points,
$\lambda$ were chosen to match corresponding quantities in the the real
catalogue. In our case $V = 3.58\cdot10^{10}$\,Mpc$^3$, and $\lambda =
5.15\cdot10^{-7}$\,Mpc$^{-3}$.  The total number of points within $V$,
$N=V\lambda \approx 18400$. To account for the edge effects the mock
catalogues were generated over the larger area stretching out $250$\,Mpc
around $V$.

The solid curve gives the expected number of centres according to the
analytic formula \citep{soltan85}:

\begin{equation}
n(R) = V \lambda \,e^{-\frac{4}{3} \upi\,R^3 \lambda}\,
     \left(\frac{\upi^4 R^6 \lambda^2}{6} - 4\,\upi\,R^3\,\lambda + 1\right)\,,
\label{eq:vd_poisson}
\end{equation}

\ni for parameters $V$ and $\lambda$ given above.
Equation~\ref{eq:vd_poisson} was derived under the assumption that for
given $R$, the void centres are not considerably larger than the sphere of
radius $R$. Thus, the formula breaks down at radii just a couple percent
greater than the percolation radius, $R_{\rm p}$, i.e. the radius at which
the individual void centres merge and create an infinite web. In the
present case $R_{\rm p} \approx 110$\,Mpc.

Even small variations of the local average matter density have a
substantial impact on space concentration of voids in the relevant range of
radii. To illustrate a steep dependence of the void probability on the
quasar density in the limit of large voids, we use the
Eq.~\ref{eq:vd_poisson}.  The dotted curve in Fig.~\ref{fig:sim_vd_number}
shows the expected number of voids in the same volume for $10$\,percent
lower density, i.e. for $\lambda = 4.64\cdot10^{-7}$\,Mpc$^{-3}$. At radius
$R \approx 171$\,Mpc the number of voids as compared to the original
$\lambda$ is higher by a factor of $2$ and the ratio increases at
greater $R$.

\section[]{Voids -- numerical algorithm}
\label{app:numerics}

Spatial resolution of all the computations is set to $1$\,Mpc. To search
the whole volume of $\sim\!3.6\cdot 10^{10}$\,Mpc for the void centres, a
several step procedure is applied. First, the volume is divided into
$\sim\!8.7\cdot 10^6$ cubic `domains' of $a = 16$\,Mpc a side. For given
void radius $R$, a distance $d$ between the domain centre and each quasar
is examined. If $d < R - a\sqrt{3} / 2$, the domain is eliminated from
further computations.  Number of accepted domains, $n_{\rm ad}(R)$, rises
rapidly with the decreasing $R$.  So, for $R$ equal to $193$, $190$, $155$,
and $145$\,Mpc the corresponding numbers of domains are: $1$, $35$,
$13544$, and $45720$. Then, space distribution of accepted domains is
inspected, and domains are segregated into clusters, where a cluster is
defined as  a group of mutually contiguous domains. Each cluster hosts one
or more void centres.

In the next step, each accepted domain is divided into $16^3$ $1$\,Mpc
`cells'.  For the each cell the distance from the cell centre to the
nearest quasar is determined. If it is smaller than $R$, the cell is
removed.  Obviously, the total number of saved cells raises with the
decreasing $R$ proportionally to $n_{\rm ad}(R)$, and quickly becomes
unmanageable. To facilitate computations over the entire interesting range
of $R$, only the cells distributed on the surface of the void centre are
saved for further processing.  A removal of the `interior' cells allows for
the effective analysis of void structures, although, the numbers of kept
cells are still quite big. The largest void found at radius $R = 145$\,Mpc
contains more that $400\,000$ `surface cells'. Space arrangement of cells
determines the distribution and shape of void centres.  Cells are split
into clumps of `neighbours'. An isolated clump of cells defines a single
void centre.  Two cells are assumed to be neighbours if their separation in
each coordinate does not exceed $2$\,Mpc.  This particular separation was
chosen to balance a discrete character of cell selection, despite the fact
that the void centre is a continuous body.

\ni Very good agreement between the number of voids detected in simulations
and given by the Eq.~\ref{fig:sim_vd_number} assures us that the present
void finding algorithm not only generates correct void numbers, but
delivers accurate information on all the remaining void characteristics, as
shape details, mutual relationships and orientation relative to the
coordinate system.

\label{lastpage}

\end{document}